\documentclass[epj]{svjour}
\usepackage{graphics}
\usepackage{wrapft}
\usepackage{lscape}
\usepackage{amsmath}
\usepackage{amssymb}
\usepackage{bm}
\usepackage{here}

\begin{document}

\title{Combined chiral and diquark fluctuations along QCD critical line and enhanced baryon production with
parity doubling}

\author{Zhao~Zhang \inst{1} \and Teiji~Kunihiro \inst{2}}
\institute{ School of Mathematics and Physics, North China Electric Power University, Beijing 102206, China \and
Department of Physics, Kyoto University, Kyoto 606-8502, Japan}
\date{Received: date / Revised version: date}

\abstract{
We argue that there should exist the large combined fluctuations of chiral and diquark condensates along
the phase boundary of QCD at moderately high density and relatively low temperature. Such fluctuations might lead to
anomalous production of nucleons and its parity partner, which we propose to detect at NICA.
}

\authorrunning{Zhang and Kunihiro}
\maketitle

%%%%%%%%%%   INTRODUCTION   %%%%%%%%%%

\section{ Three mechanisms for the possible low-temperature critical point of QCD in the presence of color superconducting phase }
\label{intro}

Since the work by Asakawa and Yazaki \cite{Asakawa:1989bq}, it is widely
believed that there may exist a chiral critical point (CP) in the $T$-$\mu$
phase diagram of QCD. Such a point may be located at the relatively high temperature
region, which is promising to be scanned by Heavy ion collisions. On
the other hand, there {\em is} a possibility that the QCD phase diagram may have new chiral CP($'s$)
in the low temperature area due to the influence of other phases, especially of the color
superconductivity. In the literature, three mechanisms for realizing the low-temperature CP($'s$)
are known, which are all due to the interplay between the chiral and diquark condensates.

1.~Vector interaction:\,
In \cite{Kitazawa:2002bc}, it is found that the vector interaction can effectively enhance
the competition between the chiral and diquark condensates and even lead to a low-temperature CP;
the competition results in a small coexistent region (COE) with both chiral and (2CSC-)diquark
condensates, and the chiral transition becomes a crossover in the low-temperature region including
zero $T$.

2. Axial anomaly in the presence of the color flavor locking (CFL) phase:\,
The possible cubic coupling between chiral and diquark condensates due to the axial anomaly
may lead to a low-temperature CP for the three-flavor case, according to the idealized analysis
based on a Ginzburg-Landau(G-L) action in the chiral limit\cite{Hatsuda:2006ps}. It is to be noted
that the color superconducting phase is necessarily the CFL in such an idealized situation, and
it is totally obscure whether the new CP thus obtained is robust in the realistic situation with
finite and different quark masses for three flavors. One should also note that the G-L analysis
would lose its validity if a phase transition is a strong first-order one.

\begin{figure}[!t]
\begin{minipage}[t]{0.4\textwidth}
\resizebox{1.0\textwidth}{!}{%
\includegraphics{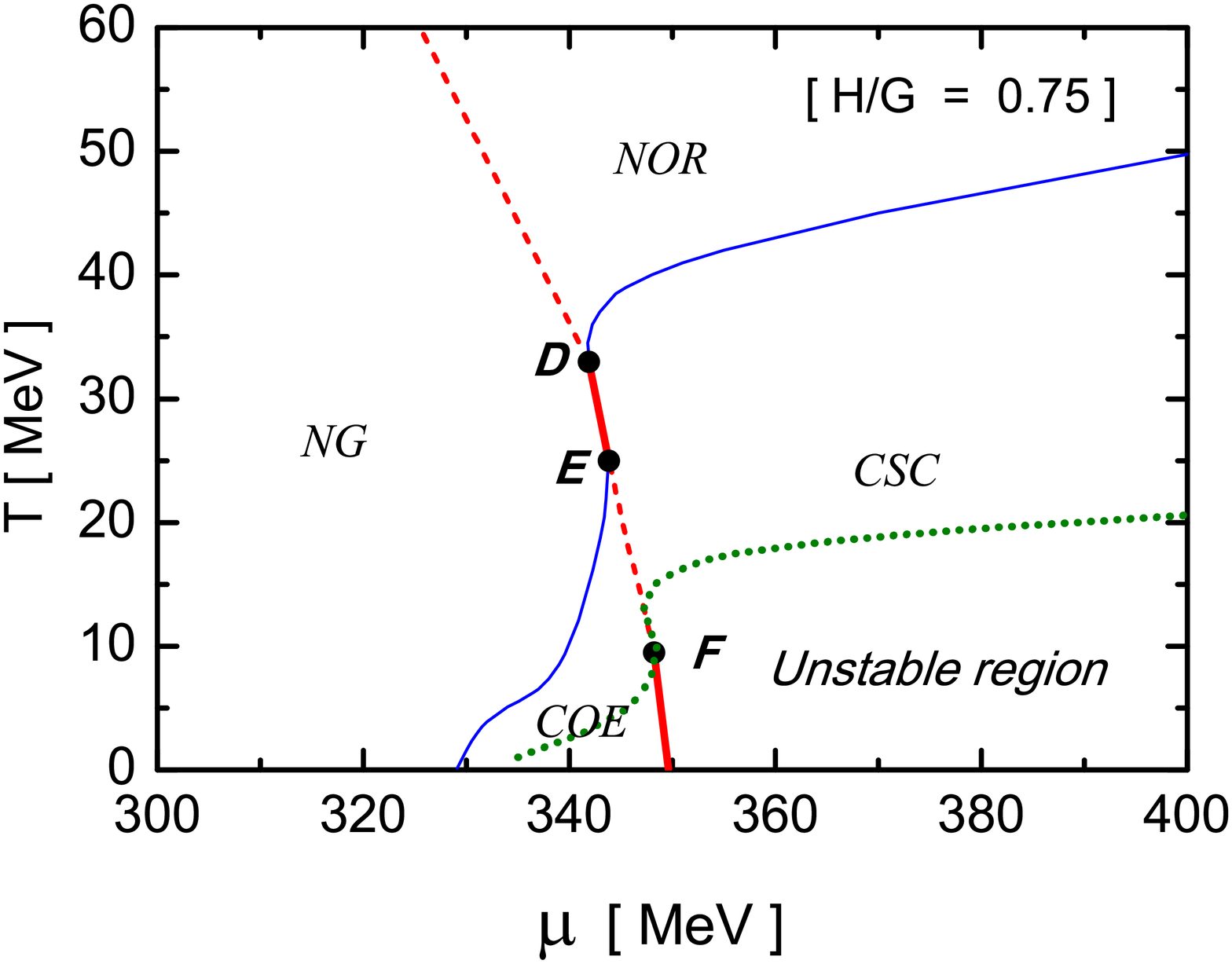}
}
\centerline{(a)}
\end{minipage}
\begin{minipage}[t]{0.4\textwidth}
\resizebox{1.0\textwidth}{!}{%
\includegraphics{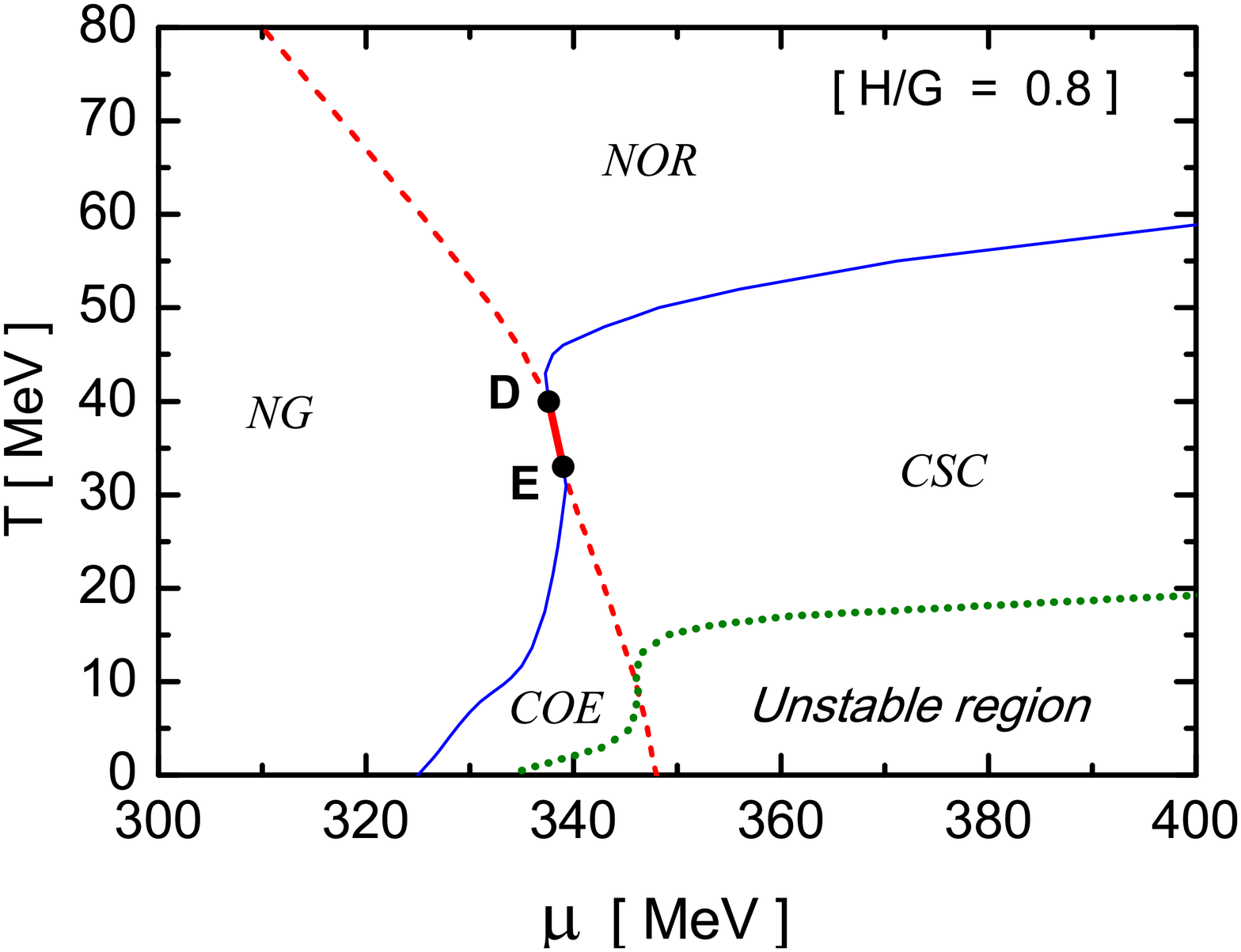}
}
\centerline{(b)}
\end{minipage}
\caption{
The phase diagram with the low-temperature CP($'s$) induced
by the electric chemical potential required by charge-neutrality in the presence of
2CSC. The thick solid line, thin solid line and dashed line denote the
first order transition, second order transition and chiral crossover, respectively. The
unstable region is characterized by the chromomagnetic instibility.
The figures are taken from \cite{Zhang:2008wx}, which are obtained in
a two-flavor NJL model.
}
\label{fig:charge}
\end{figure}

3. Electric chemical potential required by the charge-neutrality and $\beta$-equilibrium:\,
It is first reported in \cite{Zhang:2008wx} that the electric chemical potential $\mu_e$
required by the charge-neutrality can effectively strengthen the chiral-diquark interplay
and gives rise to one or even two low-temperature CP($'s$): see Fig.\ref{fig:charge}. In this
mechanism, the $\mu_e$ plays double roles on the phase transition: First, it delays the chiral
transition towards to a larger chemical potential, just like what the vector interaction does.
Second, the finite $\mu_e$ implies a Fermi-surface mismatch between u and d quarks leading to
an abnormal temperature dependence of the diquark condensate, which causes the multiple CP$'s$.

Note that for the mechanisms 1 and 3,  the 2CSC is the favored color superconducting phase
in the COE. So it is naturally expected that the COE should be enlarged when both the vector
interaction and the charge-neutrality are taken into account. This has been confirmed in \cite{Zhang:2009mk}
and even four CP$'s$ are observed in the calculation based on the NJL model.

\section{The three-flavor (two-plus-one-flavor) NJL model doesn't support the phase diagram with
a low-temperature critical point induced by the Axial Anomaly }

Since the G-L theory is solely based on the symmetry properties of the critical
point, its analysis can be model-independent once the coefficients of the terms appearing
in the action are given. However, the G-L theory itself has no ability to determine the very coefficients
and hence useless for a quantitative prediction. So a natural question is that whether such a prediction of
a new low-temperature CP based on this method could be fulfilled in reality in
QCD.  A work toward this problem has been done by Abuki et al. \cite{Abuki:2010jq} using a
three-flavor NJL model with a new axial anomaly term
\cite{Rapp:1999qa,Steiner:2005jm}
\begin{eqnarray}
\mathcal{L}_{\chi{d}}^{(6)}&=&\frac{K'}{8}\sum_{i,j,k=1}^3\sum_{\pm}
[({\psi}t_i^{f}t^{c}_k(1 \pm\gamma_5){\psi}_C)\nonumber\\
&&\times (\bar{\psi}t_j^{f}t^{c}_k(1 \pm \gamma_5)\bar{\psi}_C)
(\bar{\psi}_i( 1 \pm \gamma_5)\psi_j)\label{eqn:Lagrangian4}],
\end{eqnarray}
as well as the standard Kobayashi-Maskawa-'t Hooft interaction
\cite{Kobayashi:1970ji,'t Hooft:1976fv,'t Hooft:1986}
\begin{equation}
\mathcal{L}_{\chi}^{(6)}=-K\det_{f}\left[ \bar \psi ( 1 + \gamma_5 ) \psi \right] + {\textrm h.c.}
\end{equation}
It was found that the low-temperature CP induced by the axial anomaly is really observed again in
such an idealized model.

However, there exists two shortcomings in their study. First, as stated above,
the SU(3) flavor symmetry is assumed with $m_u(m_d)=m_s$.
Second, the new CP only appears for very strong $K'$ region
( namely, for $K'/K>K'_c/K$=3.8, where $K$ is fixed as it's vacuum value ).
It is known that the large difference between $m_s$ and $m_u(m_d)$
gives rise to a so significant effect on the phase transitions at moderate
density region that the CFL is disfavored \cite{Ruester:2005jc,Abuki:2005ms}.
Furthermore, $K'$ should become weaker due to the suppression of the instantons.

Actually, a more serious problem has been pointed out by Basler and Buballa \cite{Basler:2010xy}
by using the same model as in \cite{Abuki:2010jq}: if $K'$ is too strong, the 2CSC is favored
over the CFL assumed in \cite{Abuki:2010jq} near the chiral boundary, even in the
three-flavor symmetry case. For physical quark masses, we can expect that the CFL will be more
unlikely in competition with the 2CSC in this model because of the mass mismatch. This point has
also been confirmed in \cite{Basler:2010xy}. So the investigations in \cite{Basler:2010xy} indicate
that the NJL model does not support the proposed new CP induced by the axial anomaly
in \cite{Hatsuda:2006ps} for both the three-flavor and two-plus-one-flavor cases.

\section{Does the two-plus-one-flavor NJL model support the phase diagram with a low-temperature critical point when taking into account the Axial Anomaly, Vector Interaction, and/or Charge-neutrality?  }

Then, a question naturally arises: does the NJL model support the low-temperature
CP when taking into account the axial anomaly, the vector interaction, and the
charge-neutrality and $\beta$-equilibrium in the presence of 2CSC? The present authors
have investigated this problem in a two-plus-one-flavor NJL model by taking into account
all these ingredients\cite{Zhang:2011xi}. Our study suggests that, besides the vector
interaction $G_V$ and the electric chemical potential $\mu_e$, the new axial anomaly
interaction $K'$ also favors the COE with 2CSC. The reason is that, near the chiral boundary,
the strange quark condensate $\sigma_3$ has still a relatively large value which can effectively
enhance the Majarona mass for the u-d pairing due to the flavor mixing induced by the anomaly
term $K'$:
\begin{equation}
\Delta_3=2(G_D-\frac{K'}{4}\sigma_3)s_3\,,\label{eqn:mmass}
\end{equation}
where $s_3$ stands for the u-d diquark condensate.
So unlike its role in the mechanism 2, the heavy strange quark gives a positive contribution to
the emergence of the low-temperature CP by enhancing the chiral-diquark interplay for u and d quarks.
\begin{figure*}[!t]
%\hspace{.05\textwidth}
\begin{minipage}[t]{.42\textwidth}
\resizebox{1.0\textwidth}{!}{%
\includegraphics{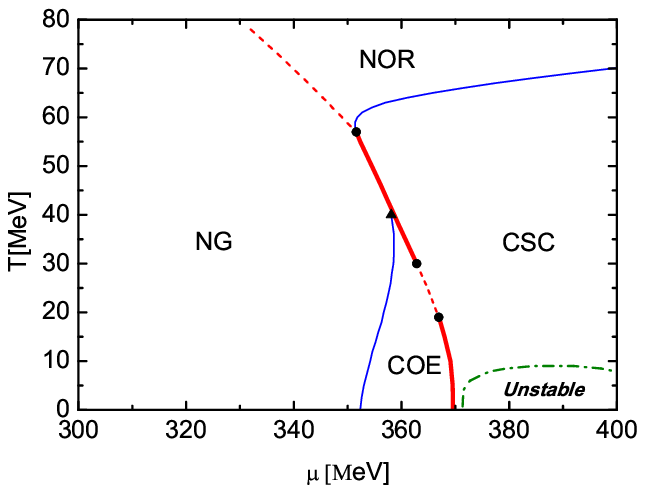}
}
\centerline{(a) $K'/K=0.55$}
\end{minipage}
%\hspace{-.15\textwidth}
\begin{minipage}[t]{.42\textwidth}
\resizebox{1.0\textwidth}{!}{%
\includegraphics{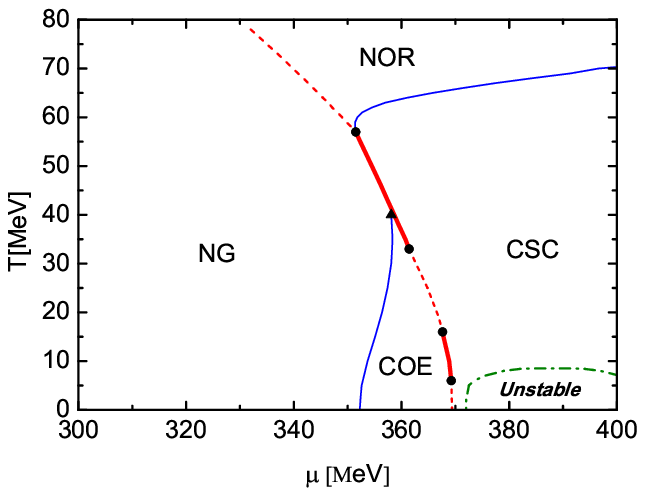}
}
\centerline{(b) $K'/K=0.57$}
\end{minipage}
%\hspace{.4\textwidth}
\begin{minipage}[t]{.42\textwidth}
\resizebox{1.0\textwidth}{!}{%
\includegraphics{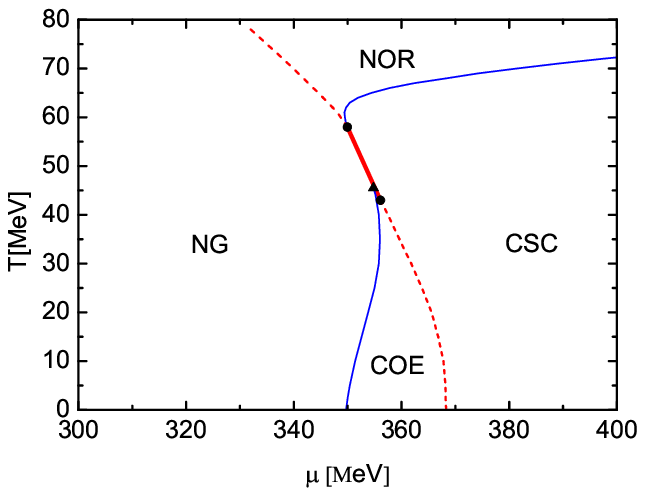}
}
\centerline{(c) $K'/K=0.70$}
\end{minipage}
\hspace{.14\textwidth}
\begin{minipage}[t]{.42\textwidth}
\resizebox{1.0\textwidth}{!}{%
\includegraphics{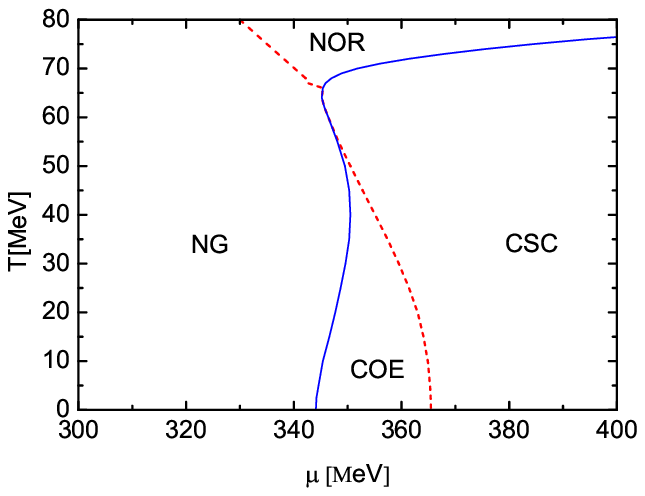}
}
\centerline{(d) $K'/K=1.0$}
\end{minipage}
\caption{The $T$-$\mu$ phase diagrams of the two-plus-one-flavor NJL model
for several values of $K'/K$ and fixed $G_V/G_S=0.25$, where
the charge-neutrality constraint and $\beta$-equilibrium condition
are imposed. The meanings of the different line types are the same as those in Fig.~\ref{fig:charge}.
With the increase of $K'/K$, the number of the
critical points changes and the unstable region characterized
by the chromomagnetic instability (bordered by the dash dotted line)
tends to shrink and ultimately vanishes in the phase diagram. The figures
are taken from \cite{Zhang:2011xi}.
}\label{fig:pdGVfixed}
\end{figure*}

\begin{figure*}[!t]
%\hspace{-.0\textwidth}
\begin{minipage}[t]{.33\textwidth}
\resizebox{1.0\textwidth}{!}{%
\includegraphics{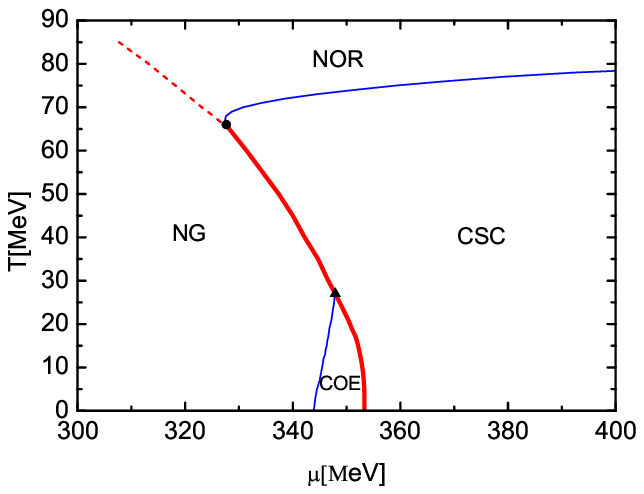}
}
\centerline{(a) $G_V/G_S=0$}
\end{minipage}
%\hspace{-.05\textwidth}
\begin{minipage}[t]{.33\textwidth}
\resizebox{1.0\textwidth}{!}{%
\includegraphics{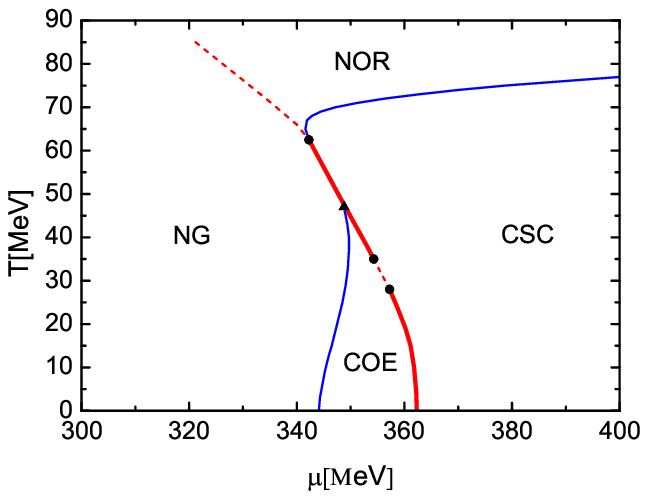}
}
\centerline{(b) $G_V/G_S=0.193$}
\end{minipage}
\begin{minipage}[t]{.33\textwidth}
\resizebox{1.00\textwidth}{!}{%
\includegraphics{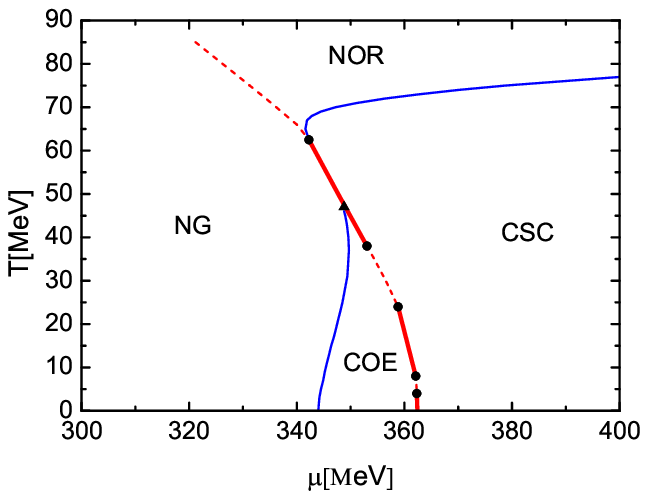}
}
\centerline{(c) $G_V/G_S=0.195$}
\end{minipage}
\begin{minipage}[t]{.33\textwidth}
\resizebox{1.0\textwidth}{!}{%
\includegraphics{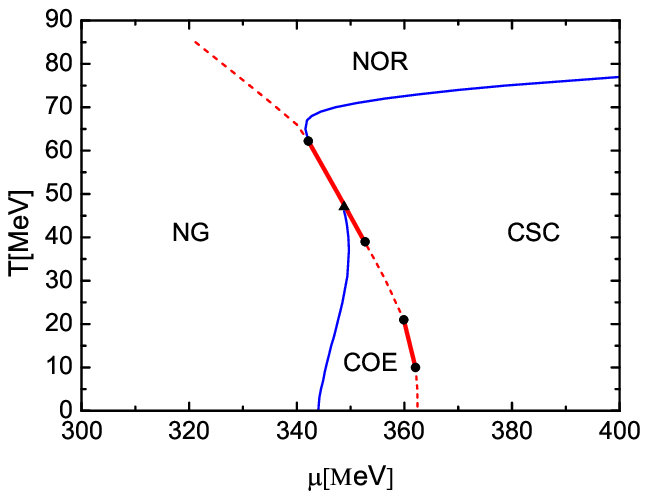}
}
\centerline{(d) $G_V/G_S=0.197$}
\end{minipage}
\begin{minipage}[t]{.33\textwidth}
\resizebox{1.0\textwidth}{!}{%
\includegraphics{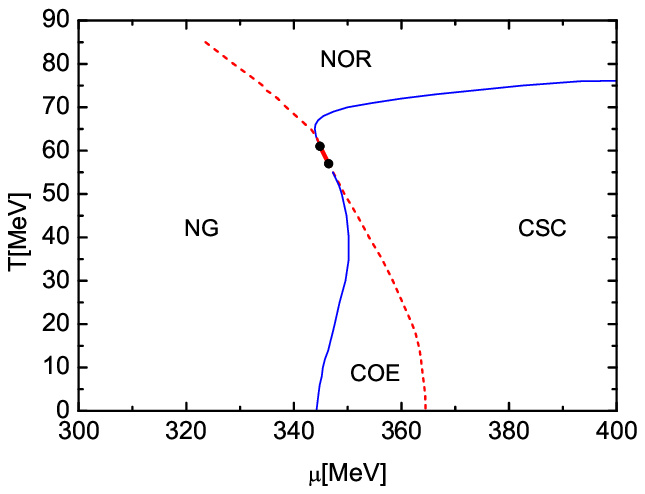}
}
\centerline{(e) $G_V/G_S=0.23$}
\end{minipage}
\begin{minipage}[t]{.33\textwidth}
\resizebox{1.0\textwidth}{!}{%
\includegraphics{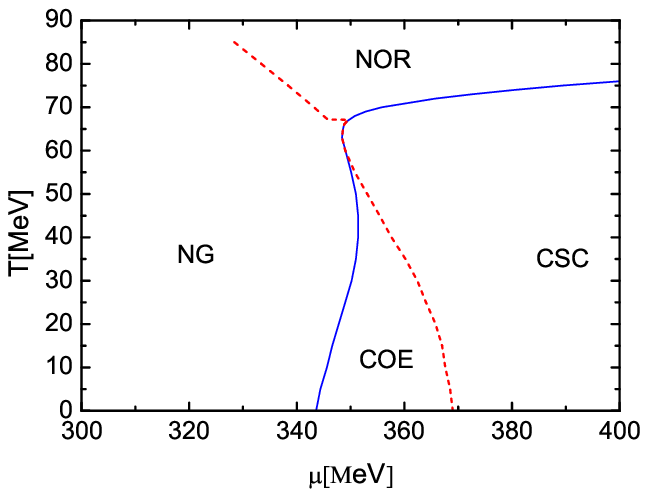}
}
\centerline{(f) $G_V/G_S=0.3$}
\end{minipage}
\caption{
The $T$-$\mu$ phase diagrams in the two-plus-one-flavor NJL model for
fixed $K'/K=1.0$ with $G_V/G_S$ being varied, where the charge-neutrality constraint
and $\beta$-equilibrium condition are taken into account. The meanings
of the different line types are the same as those in Fig.~\ref{fig:charge}.
The number of the critical points changes along with an increase of $G_V/G_S$. All
the phase diagrams are free from the chromomagnetic instability. The figures are
taken from \cite{Zhang:2011xi}.}
\label{fig:pdK'fixed}
\end{figure*}

Figure.~\ref{fig:pdGVfixed} shows the $T$-$\mu$  phase diagram of a two-plus-one-flavor NJL with
the charge-neutrality and $\beta$-equilibrium for different $K'$ and fixed vector interaction
$G_V$. Due to the strengthened chiral-diquark competition, the low-temperature CP($'s$) and crossover
for the chiral transition appear in the $T$-$\mu$ plane. One observes that the number of the critical
points changes as 1$\rightarrow$\, 3\,$\rightarrow$\,4\,$\rightarrow$\,2\,$\rightarrow$\,0 when
$K'$ is increased. Owing to the effect of $G_V$ and $\mu_e$,  the low-temperature CP($'s$) can be
realized with a relatively small $K'$  ( Note that without the charge-neutrality constraint,  we do
not find the low-temperature CP in this model if only the axial anomaly and vector interaction are
considered ). In Fig.~\ref{fig:pdK'fixed}, the similar chiral CP structures appear for fixed $K'/K=1$
(The Fierz transition of Kobayashi-Maskawa-'t Hooft interaction gives $K'/K=1$ ) but varied $G_V$.

Thus one sees that the chiral-diquark interplay in the COE region becomes complicated once
the axial anomaly, the vector interaction and the charge-neutrality are all taken into account.
Figures.~\ref{fig:pdGVfixed} and \ref{fig:pdK'fixed} tell us that there exists a parameter region
in the $K'$-$G_V$ plane for the low-temperature CP($'s$) in the NJL model. The above parameter
area seems to include the physical region since $K'$ is expected to be suppressed near the chiral
boundary while the instanton molecular liquid model predicts that $G_V/G_S$=0.25.
.
\begin{figure}[!t]
\begin{minipage}[t]{.43\textwidth}
\resizebox{1.0\textwidth}{!}{%
\includegraphics{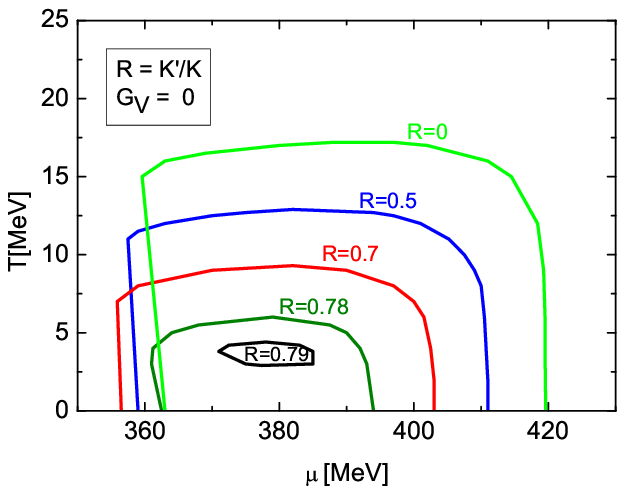}
}
\centerline{(a)}
\end{minipage}
\begin{minipage}[t]{.43\textwidth}
\resizebox{1.0\textwidth}{!}{%
\includegraphics{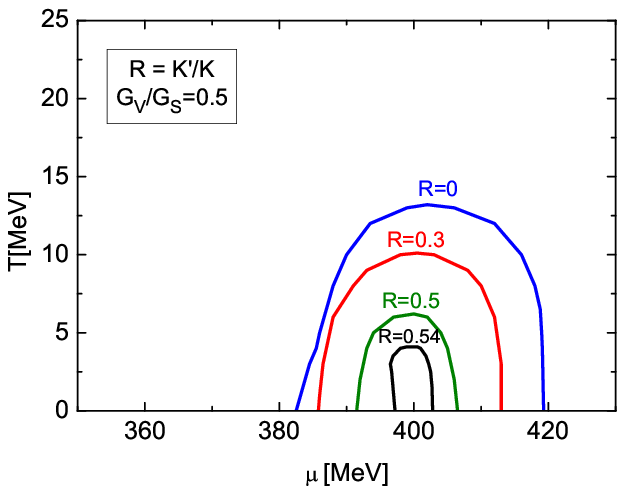}
}
\centerline{(b)}
\end{minipage}
\caption{
The boundary between the stable and unstable homogenous 2CSC
regions with (upper figure) and without (lower figure) the vector
interaction in two-plus-one-flavor NJL model. With the increase of
the ratio $K'/K\equiv R$, the unstable region with the chromomagnetic
instability in the $T$-$\mu$ plane shrinks and eventually vanishes.
The figures are taken from \cite{Zhang:2011xi}.}
\label{fig:unstable-stable}
\end{figure}

\section{The suppression of Chromomagnetic Instability by Axial Anomaly and Vector Interaction  }
\label{sec: instability }

Another important role of the axial anomaly and vector interaction is that they can effectively
suppress the chromomagnetic instability \cite{Huang:2004bg} associated to the gapless
2CSC \cite{Shovkovy:2004me}. This is shown in Fig.~\ref{fig:unstable-stable}: The unstable region
with the instability in the $T$-$\mu$ plane shrinks with increasing $K'$. Figure.~\ref{fig:unstable-stable}
also tells us that such a suppression becomes more significant when the vector interaction is included.

The reason for the suppression of the chromomagnetic instability can be attributed to
two facts. First, the u-d diquark coupling is enhanced by the s quark due to the nonzero
$K'$, which can suppress the instability \cite{Kitazawa:2006zp}. Second, the effective chemical
potential difference is shifted towards smaller values by $G_V$. So the hot asymmetric homogeneous
2CSC may be free from the instability and thus could be formed at NICA densities.

\section{Implication for the experiments at NICA }

Our study suggests that the mixed phase with both chiral symmetry breaking and 2CSC
may coexist in the low temperature and moderate density region near the chiral boundary.
The interplay of chiral and diquark condensates can be enhanced by the axial anomaly,
vector interaction, and electric chemical potential, which may lead to low temperature
chiral crossover and CP($'s$). For not very low-temperature, such a color superconducting
phase may be free from the chromomagnetic instability. This indicates that the mixing phase
with 2CSC may appear in the neutron star mergers or core-collapse supernova explosions, where
the temperatures can reach several tens of MeV.

It would be of course interesting if this possible phase structure can be explored
in the future experiments of heavy ion collision at NICA, which aiming at the physics of
intermediate baryon density of QCD. Before applying the above results to heavy-ion collisions, we
should note that although the charge neutrality as described by $\mu_e$ plays an important role
in the above analysis, the charge neutrality is not satisfied in heavy-ion collisions.
However, it can be partially replaced by the effect of a finite isospin chemical potential
present in heavy-ion collisions, which implies that the possible low-temperature crossover or CP
of our findings can be tested at NICA.

In applying our result to heavy-ion collisions, we should first mention that our results are obtained
in the mean-field approximation without including quantum and thermal fluctuations which should play a
significant role in the vicinity of the critical point. Thus the appearance of the multiple critical points
may suggest that there exist large fluctuations of the chiral and diquark fields around the phase boundary
in reality \cite{Kunihiro:2010vh}. In such a region of the $T$-$\mu$ plane, the chiral symmetry is partially
restored and simultaneously the diquark correlations are strong, which may imply that there exists a
significant amount of preformed diquarks \cite{Kitazawa:2001ft,Kitazawa:2003cs,Kitazawa:2005vr}.
Furthermore, if the baryon number density is relatively low, we can expect the BCS-BEC crossover
\cite{Shuryak:2004tx,Nishida:2005ds,Sun:2007fc,Kitazawa:2007zs,Ferrer:2014ywa} toward lower density where the
inter-quark interaction is stronger while the mean inter-quark distance is smaller and hence the formation
of bound diquark states in the configuration space is expected. On the basis of this observation, we can
speculate that the proliferation of the baryons may happen in the final state of heavy-ion collision if the
system enters this phase. This is because the preformed diquarks provide the seeds of nucleons. Moreover, since
the chiral symmetry is partially restored, the partial parity doubling may occur in the baryon sector.
Thus, we propose that the enhanced production of the nucleon and its parity partner N$^*$(1535) can serve as the
experimental signal of the hot (partially) chiral symmetric CSC quark matter which may be created in NICA.

Finally, a few comments are in order:
1)~We have suggested that the possible realization of a (partially) chiral symmetric phase may be detected as
parity doubling in the baryon sector. In this respect, it is worth mentioning that chirally symmetric but confined
excitations are also discussed in the context of quarkyonic matter in \cite{McLerran:2007qj}, where the argument
is based on the large $N_c$ counting.
2)~In our study, we have taken it for granted that the deconfinement transition happens before
the appearance of the COE. We mention that such a deconfined phase with massive quarks
seems to be supported also by the percolation analysis in \cite{Castorina:2010gy},
where the color superconductivity is not explicitly considered, though.
3)~Recently, people are interested in the inhomogeneous chrial condensate
to be realized at moderately high densities \cite{Buballa:2014tba}.
Certainly this is an interesting topic and may have a relevance to the phenomenology of compact stars.
However, it might be unlikely that such crystal-like phases can survive, owing to the thermal fluctuations of mesonic
fields \cite{Lee:2015bva,Hidaka:2015xza} in the collisions of heavy-ions with a finite geometry.

\section*{Acknowledgment}
Z.Z. was supported by the National Natural Science Foundation of China ( No.11275069 ). T.K. was partially supported by a
Grant-in-Aid for Scientific Research by the Ministry of Education, Culture, Sports, Science and Technology (MEXT) of
Japan ( No.20540265 ), by Yukawa International Program for Quark-Hadron Sciences, and by the Grant-in-Aid for the globl
COE program `` The Next Generation of Physics, Spun from Universality and Emergence '' from MEXT.

\end{document}